\documentstyle[preprint,aps]{revtex}
\tighten
\begin{document}
\widetext
\bigskip
\vspace{0.2in}
\thispagestyle{empty}
\begin{flushright}
NYU-TH/99/04/02, HUTP-99/028, NUB 3202, EFI-99-19\\
November 4, 1999
\end{flushright}

\vspace{0.2in}

\begin{center}
\bigskip\bigskip
{\large \bf Zerobrane Matrix Mechanics, Monopoles  
and Membrane Approach in QCD}
\vspace{0.3in}      

{Gregory Gabadadze$^1$\footnote{E-mail: gabadadze@physics.nyu.edu} 
and Zurab Kakushadze$^{2,3,4}$\footnote{E-mail: zurab@string.harvard.edu} }
\vspace{0.2in}

{\baselineskip=14pt \it 
$^1$Department of Physics, 
New York University, New York, NY 10003  \\
$^2$Jefferson Laboratory of Physics, Harvard University, 
Cambridge, MA  02138 \\
$^3$Department of Physics, Northeastern University, Boston, 
MA 02115\\
$^4$Enrico Fermi Institute, University of Chicago, Chicago, IL 60637} \\
\vspace{0.2in}
\end{center}

\vspace{0.9cm}
\begin{center}
{\bf Abstract}
\end{center}
\vspace{0.3in}

We conjecture that a T-dual form  of 
pure QCD describes dynamics of point-like monopoles. 
T-duality transforms the QCD Lagrangian into a 
matrix quantum mechanics of zerobranes which we identify with monopoles. 
At generic points of the monopole moduli space
the $SU(N)$ gauge group is broken down
to $U(1)^{N-1}$ reproducing the key feature of 't Hooft's 
Abelian projection.
There are certain points in the moduli space
where monopole positions coincide, gauge symmetry is 
enhanced and gluons emerge as massless excitations.
We show that there is a linearly rising potential 
between zerobranes. This indicates the presence of a stretched 
flux tube between monopoles.  
The lowest energy state is achieved when monopoles are sitting on top
of each other and gauge symmetry is enhanced. 
In this case they behave as free massive
particles and can condense. In fact, we find a constant
eigenfunction of the corresponding Hamiltonian
which describes condensation of monopoles. 
Using the monopole quantum mechanics, 
we argue that large $N$ QCD in this T-dual picture is a 
theory of a closed bosonic membrane
propagating in {\em five} dimensional space-time. QCD   
point-like monopoles can be regarded  in this approach 
as constituents of the membrane.

\newpage
\noindent
{\bf 1. Introduction and Summary}
\vspace{0.1in} 

The aim of this work is to study point-like monopoles
in pure $SU(N)$ QCD (Yang-Mills (YM) theory). 
There is a great amount of evidence, both theoretical and from the 
lattice, that this theory confines colored charges. 
It is conjectured  that confinement is realized as a 
dual Meissner effect of  superconductivity 
\cite{tHooft1,Mandelstam,Nambu}\footnote{More precisely, 
one is talking about
S-duality here. For   
recent discussions of these issues, see, {\em e.g.}, 
\cite {Brambilla,Backer}.}. 
In superconductivity fundamental 
charges are condensed in Cooper pairs and 
monopoles can be confined by vortices \cite{Abrikosov}. 
Therefore, in the dual picture  which is believed to be QCD,  
flux tubes should be connecting colored charges and  
the corresponding magnetic monopoles must be condensed.
Thus, a crucial step in studying confinement is to
understand  monopole dynamics.
Great progress in this direction was made 
by Seiberg and Witten within the framework of 
supersymmetric counterparts of QCD \cite{SeibergWitten,Seiberg},
where monopole condensation and dual superconductivity picture
were explicitly shown to be realized. 
However, pure $SU(N)$ QCD, as opposed to 
its supersymmetric counterparts, lacks any scalar fields. Therefore, 
there are no 't Hooft-Polyakov type of monopoles 
in the model. Moreover, the strong coupling dynamics
makes it difficult to find some other type of solutions.   
Nevertheless, it was 't Hooft \cite{tHooft2} who argued  that 
QCD monopoles  could be hidden behind the redundancy
of the gauge invariant description of the theory.
Thus, according to 't Hooft \cite{tHooft2},
one can choose  a certain class of unitary gauges 
in which there are no propagating spurious fields,
and the non-Abelian part of the gauge freedom is 
completely fixed. In these gauges  the $SU(N)$ 
gauge symmetry is broken down to its maximal Abelian
subgroup $U(1)^{N-1}$. Each Abelian
subgroup gives rise to monopoles. 
Thus, there should exist  $N-1$ different 
types of point-like monopoles in QCD (for a recent review, see 
\cite{Polikarpov}). 

The way these monopoles  can actually be seen in the theory is
a bit peculiar. Unlike 't Hooft-Polyakov
monopoles, they emerge somewhat indirectly, in particular,  
as point-like singularities occurring for certain 
gauge choices \cite {tHooft2}. This makes it difficult 
to study their dynamics. In this respect, the main 
questions arising in this approach are the following:

\begin{itemize}

\item {Can QCD monopoles and their interactions  be studied  
within the conventional Hamiltonian approach?}

\item {Can monopole condensation be analytically studied 
using dynamics of the QCD monopoles?}

\item {Can the gauge group breaking pattern   
$SU(N) \rightarrow U(1)^{N-1}$ be understood without  referring
to any specific gauge conditions, but rather relying on 
underlying dynamics of the theory?}

\end{itemize}

In the following we will argue that, under certain assumptions,
the answer to these questions can be  {\it positive}.

In order to discuss these issues, one needs the notion of
S- and T-duality. Under S-duality, the roles of fundamental 
quanta and solitons are interchanged \cite {MontonenOlive}.   
As we discussed above, QCD might  be  S-dual to 
some model of superconductivity, and  condensation of
QCD monopoles should indicate confinement of QCD colored charges
\cite {tHooft1,Mandelstam}. However, as we have also 
emphasized above, it is very difficult to study  monopoles within the 
strongly coupled Lagrangian of QCD. Therefore, some other approach is 
called for. 
T-duality, which in general interchanges momentum modes with 
winding modes,  seems to be a promising approach in this case. 
Below we will  argue that strongly coupled  
pure $SU(N)$ QCD in a large but finite volume 
can be described in a T-dual form 
as a certain matrix quantum mechanics of $N-1$ 
point-like zerobranes. We conjecture that the zerobranes in the T-dual picture 
can be identified with QCD monopoles.   
We use the  matrix quantum mechanics to study 
certain properties of QCD zerobranes (monopoles). 
In fact, we show that the $SU(N)$
gauge group is  
generically broken down to $U(1)^{N-1}$ as a result of 
zerobrane  dynamics. Furthermore,
we show that in a certain approximation 
(the so called monopole moduli space approximation) 
of heavy, slowly-moving, almost non-interacting  monopoles 
matrix quantum mechanics has the unique ground state 
with zero momentum. This describes monopole condensation.   

In order to reveal the structure of the condensate
and validity of the monopole moduli space approximation
we study monopole interactions. In fact, we find a 
linearly rising potential between them. 
Thus, there should be flux tubes stretched between monopoles. 
As a result of this observation 
we conclude that the moduli space approximation
mentioned above turns out to be valid only when 
monopoles are placed on top of each other. 
In this case there is no force between them 
and they will  be condensing at these points of the moduli space.

There is one more crucial feature associated with 
the points of the moduli space where monopoles
sit on top of each other. We argue that at these points the
$U(1)^{N-1}$ 
gauge group is enhanced back to $SU(N)$,
and, as a result, {\it massless} gluons emerge 
in the theory. To summarize, the following
attractive picture emerges. At a generic point 
of the zerobrane  moduli space monopoles are
separated and the gauge group is broken
down to $U(1)^{N-1}$. There are flux tubes
stretched between zerobranes. They give rise to linearly 
rising potential between these heavy point-like 
objects. When the monopoles come on top of each other,
they condense; moreover, in this case the broken gauge
group is enhanced back to $SU(N)$ and gluons
emerge as massless states. 

Finally, using the T-dual description mentioned above, 
we show that in the large $N$ limit pure QCD
can be described as a $(4+1)$-dimensional theory of {\it a closed 
bosonic membrane}. It is tempting to conjecture that perhaps this theory
in a certain approximation/limit
could be thought of as a non-critical closed bosonic  string theory
which is believed to describe large $N$ QCD \cite{Polyakov}. 

The rest of this paper is organized as follows. In section 2 we discuss 
T-duality of pure QCD. In section 3 we study quantum mechanics 
of zerobranes. We show that there is a linearly 
rising potential between these point-like objects.
In section 4 we discuss condensation of QCD monopoles.
In section 5 we show that the large $N$ limit of 
pure QCD can be related via T-duality  to a theory of a
$(4+1)$-dimensional bosonic  membrane. 
In section 6 we speculate on a possibility of an underlying 
five dimensional theory which might provide an adequate low energy 
description of strongly coupled  QCD. 

\newpage
\noindent
{\bf 2. Pure QCD on a Torus}
\vspace{0.1in}

In this section we deduce the action 
describing pure QCD on a spatial torus. 
Consider pure $SU(N)$ QCD with the Lagrangian density:
\begin{eqnarray}
 {\cal L}_{\rm YM}~=~-{1\over 4 g^2_{\rm YM}}~G_{\mu\nu}^a
 G^{a\mu\nu}~.
\label{YMlagrangian}
\end{eqnarray} 
To avoid complications with the Gribov copies, 
we will be working in the $A_0=0$ gauge. 
The Hamiltonian density of the model can be written as follows:
\begin{eqnarray}
 {\cal H}_{\rm YM}~=~{ g_{\rm YM}^2\over 2}~P_i^2~+~
   {1\over 4 g^2_{\rm YM}}~G_{ij}^2~,
\label{hamiltonian}
\end{eqnarray}
where $i,j=1,2,3$, and  the canonically conjugate momenta are
defined as  $g^2_{\rm YM}P_i^a=-G_{0i}^a$. The Gauss's law
should  be imposed on physical 
eigenstates of the Hamiltonian:
\begin{eqnarray}
(D_i~P_i)^a~|{\rm Phys.}\rangle~=~0~.
\label{gauss}
\end{eqnarray}
The theory is known to generate the mass scale  
$\Lambda_{\rm YM}$. The corresponding effective correlation 
length will be defined  as follows:
\begin{eqnarray}
\zeta~\equiv~ {1\over \Lambda_{\rm YM}}~.
\label{zeta}
\end{eqnarray}
Let us assume that the theory is placed in a 
finite volume $V$ (we will specify this volume and boundary 
conditions below). It is useful to introduce the following two
limits. One can define the value of the volume element 
$V$ to be small if the correlation length $\zeta$ is much larger than  
$V^{1/3}$, {\em i.e.},  $\zeta >>V^{1/3}$ \cite {Luscher}. 
The large volume limit would then 
refer to a volume element satisfying $V^{1/3}>>\zeta$.
The two limits defined above correspond to
the weak respectively strong  coupling regimes of the theory \cite
{Luscher}. 
In order to see this let us keep the product of 
the renormalization scale $\mu$ and the value of $V^{1/3}$ fixed: 
for simplicity we choose $V\mu^3=1.$  
Thus, the scale $\mu$ serves as
an infrared cutoff.
Then the expression for the correlation length
in the next-to-leading order approximation takes  the form:
\begin{eqnarray}
{\zeta \over  V^{1/3}}~\simeq~(\alpha_s)^{\beta_1\over 2 \beta_0^2}~
 \exp\left({2\pi\over \alpha_s\beta_0}\right)~.
\label{weakstrong}
\end{eqnarray}
Here $\alpha_s=\alpha_s(\mu)$ is the running strong coupling constant,
and $\beta_0$ and $\beta_1$ are the corresponding one- respectively two-loop
coefficients of 
QCD beta function. 
Using (\ref {weakstrong}),  
one finds  that
the small volume approximation, 
defined as $V^{1/3}<<\zeta$,  corresponds
to the weak coupling regime, {\em i.e.},  $\alpha_s << 1$. 
Let us now  turn to the large volume limit defined
as $V^{1/3}>>\zeta $. This  limit is equivalent to
the  small $ \mu $ approximation. Furthermore, for small values of $\mu$
the running coupling constant $\alpha_s$ is  a large number. 
In fact, for $\mu=\Lambda_{\rm YM}$ the perturbative 
strong coupling constant blows up. 
Hence, 
the large volume limit corresponds to the strong coupling regime of
the theory. In this regime the approximation (\ref {weakstrong})
breaks down\footnote{
The exact expression for $\zeta$ 
can also be given (see, for instance, \cite
{Collins}), however, this expression contains  
the exact form of the beta 
function $\beta(\alpha_s)$
which is known only perturbatively.}.
\vspace{0.2in} \\
{\it 2.1. T-dual description of pure QCD}
\vspace{0.1in}

As it was mentioned above, we 
assume that our system is placed in a finite volume $V$.
In fact we take this volume to be a 
cubic box of the size   
$2\pi L\times  2\pi  L\times  2\pi L$. 
To preserve Lorentz invariance
of the theory we impose periodic boundary conditions
on gluon fields:
\begin{eqnarray}
A_j(x_i~+~2\pi L,~t)~\equiv~A_j(x_i,~t)~.
\label{periodic}
\end{eqnarray}
(The equivalence here  is up to gauge transformations.)
This means that our 
theory is actually compactified on a three-torus  $T_{L}\equiv
S^1\times  S^1 \times  S^1$ which has  all the radii equal to $L$.

Recently it has been established that the 
(super-)Yang-Mills 
model on a torus $T_{L}$ can be rewritten in terms of a
certain matrix quantum mechanics defined on a dual torus $T_{R}=
{\widetilde S}^1\times {\widetilde S}^1 \times {\widetilde S}^1$ 
with the radii  $R$  
\cite {Danielson,KP,DKPS,BFSS,Taylor}\footnote{The original proof 
is based on the string theory arguments \cite {Danielson,KP,DKPS,BFSS}. 
However, it is applicable to 
the non-Abelian field theory case as well \cite {Taylor} if the
following identifications are made. In our interpretation, 
one can think of QCD point-like 
monopoles as  zerobranes of the original string theory picture
and the flux tubes stretched between monopoles (see below) 
as open strings. A more detailed discussion of this analogy will be given 
in the next section.}:
\begin{eqnarray}
R~=~{\alpha'\over L}~,
\label{Tdual}
\end{eqnarray}
where  the parameter $\alpha'$ is defined via  the 
QCD scale 
$$
\alpha ' ~=~ {1\over \Lambda^2_{\rm YM}}~=~\zeta^2~.
$$
Dynamical variables in the T-dual theory are
time-dependent matrices $\Phi_i(t)$, $i=1,2,3$, 
which transform in the adjoint representation:
\begin{eqnarray}
\Phi_i(t) \rightarrow U ~\Phi_i(t)~U^+~.
\label{adjoint}
\end{eqnarray}
Here $U$ stands for an $SU(N)$ matrix independent of time 
and spatial coordinates. In addition to the color indices,
for a given value of the index $i=1,2,3$, 
the matrix valued field $\Phi^i_{a,b}$, $a,b=1,...,N$, 
carries a pair of new indices corresponding to
winding modes (which, in the original picture, that is, QCD compactified on 
$T_L$, correspond to the Kaluza-Klein momenta), 
$\Phi^i_{ma,nb}$, $m,n\in{\bf Z}$ \cite {Taylor}. The
periodicity condition for
these variables on $T_R$ takes the form \cite {Taylor}:
\begin{eqnarray}
\Phi^i_{na,nb}~\rightarrow \Phi^i_{(n-1)a,(n-1)b}~+~2\pi R
 \delta_{ab}~, ~~~
\Phi^i_{na,mb}~\rightarrow \Phi^i_{na,mb}~,~~~n\neq m~.
\label{period}
\end{eqnarray} 
In what follows we will be suppressing the winding  indices
assuming that all the traces in the expressions below are taken
with respect to these indices as well, 
and that an appropriate
finite normalization of the ${\rm Tr}$ operation is
preserved by dividing these expressions 
by the corresponding  infinite factor. 
Given these conventions,
the T-dual Lagrangian can be written as 
follows \cite {DKPS,BFSS,Taylor}:
\begin{eqnarray}
{L(t)} ~=~  {1\over 2  g \sqrt{\alpha'}} \Big (
{\rm Tr}~{\dot \Phi}^2_i ~+~{1\over 2 (2\pi\alpha')^2}~{\rm Tr}~
\Big [ \Phi_i~\Phi_j \Big ]^2~\Big )~,
\label{Philagrangian}
\end{eqnarray} 
where  ${\dot \Phi}_i$ denotes the time derivative of $\Phi_i$.
The new coupling constant $g$ is related to
the pure QCD coupling as follows:
\begin{eqnarray}
{g^2_{\rm YM}\over 4\pi}~=~g~\Big ({L\over R} \Big )^{3/2}~.
\label{couplings}
\end{eqnarray}
The T-duality transformation 
which recast the original Yang-Mills 
Lagrangian density (\ref {YMlagrangian})
into the  Lagrangian (\ref {Philagrangian})
(and {\em vise-versa}) can be written as follows \cite {Taylor}:
\begin{eqnarray}
\Phi_i~\leftrightarrow ~(2\pi \alpha')~iD_i~,~~~{\dot \Phi}_i^2
\leftrightarrow (2\pi \alpha')^2~G_{0i}^2~,~~~
{\rm Tr}~\leftrightarrow~{1\over (2\pi L)^3}
\int d^3 x~{\rm tr}~.
\label{dualtransform}
\end{eqnarray}
Here ${\rm Tr}$ goes over both color and winding indices,
while ${\rm tr}$ goes over color indices only.
Our convention for the covariant derivative is
as follows: $D_j=\partial_j-iA_j$, and $G_{ij}=i[D_i~D_j]$.
One can think of $\Phi_j$ as a matrix representation
for the covariant derivative operator $D_j$ on a torus. 
Transformations for the radii and coupling constants are
given in (\ref {Tdual}) and (\ref {couplings}), respectively.
One can check that the T-duality transformation gives rise to
a T-dual form  of the Gauss's law (\ref {gauss}):
\begin{eqnarray}
f^{abc}~\Phi_i^b~\Pi_{i}^c~|{\rm Phys.}\rangle ~=0~.
\label{gauss1}
\end{eqnarray} 
Here $\Pi_{i}$ denote the momenta canonically conjugate to
$\Phi_i$, and $f^{abc}$ stand for the $SU(N)$ structure constants. 
Therefore, we see that pure QCD on a torus is described by a certain quantum 
mechanics (\ref {Philagrangian}). The variables in this Lagrangian are 
matrix-valued adjoint fields $\Phi_i$ which carry the color as well as 
winding label. In the next section we
discuss the relation of (\ref {Philagrangian}) with the compactified form
of the matrix quantum mechanics of zerobranes.  
\vspace{0.3in} \\
{\bf 3. Matrix Model}
\vspace{0.1in}

Let us consider the quantum mechanical system which is described by the 
following Lagrangian in $(3+1)$-dimensional Minkowski space-time:
\begin{eqnarray}
{L(t)} ~=~  {1\over 2  g_s \sqrt{\alpha'}} \Big (
{\rm Tr}~{\dot X}^2_i ~+~{1\over 2 (2\pi\alpha')^2}~{\rm Tr}~
\Big [ X_i~X_j \Big ]^2~\Big )~,
\label{MM}
\end{eqnarray}
where $X_i$ denote time-dependent $N\times N$ matrices. 
The Lagrangian (\ref {MM}) possesses a global $SU(N)$
symmetry:
\begin{eqnarray}
X_i(t) \rightarrow U ~X_i(t)~U^+~.
\label{Xtrans}
\end{eqnarray}
It will become evident below, that this Lagrangian describes
$N-1$ point-like massive states in (3+1)-dimensional space-time
with linearly rising potential between them.
Subsequently, we will see that after compactification on
the torus $T_R$ the Lagrangian (\ref {MM}) coincides
with that in  (\ref {Philagrangian}). This shows  the relevance of this
theory for pure QCD (in fact T-duality between them). 
Before making any contact with QCD, let us study  properties of the 
theory defined by (\ref {MM}).  
Let us start with the ground state of the model.
The potential  
\begin{eqnarray}
V~=~-~{1\over 4g_s\sqrt{\alpha'}(2\pi\alpha')^2}~{\rm Tr}~
 \Big [ X_i~X_j \Big ]^2
\label{potential0}
\end{eqnarray}
is positive definite and has flat directions. The state of
zero energy is given  by the fields for 
which  $[X_i~X_j]~=0$. 
Hence, for  the vacuum configurations all the $X_i$'s can be 
diagonalized simultaneously and take values in the Cartan subalgebra
of $SU(N)$.  These configurations can be parametrized 
in a gauge invariant way in terms of  
invariant eigenvalues of the 
adjoint matrix $X$\footnote{This invariance is up 
to the Weyl subgroup of $SU(N)$ - see below.}. 
Thus, the vacuum state can 
be described  by the following order parameters:
\begin{eqnarray}
X_i^{\rm cl}~=~\left (
\begin{tabular}{c c c c}
$r_i^1$ & 0     & $...$ & 0 \\
0     & $r_i^2$ & $...$ & 0 \\
...   & ...   & $...$ & ... \\
0     & ...   & $...$ & $r_i^N$ \\
\end{tabular}
\right )~,
\label{Phivacuum}
\end{eqnarray}
where $r^m_i$'s are the real-valued constant eigenvalues of $X_i$. 
Different values of $r^m_i$ parametrize 
different points of the vacuum manifold, that is, the moduli space. 
Generically, the constants $r^m_i$ 
are all different. When this is the case,
the global symmetry of the theory is broken down to $U(1)^{N-1}$, 
much like what happens with the local symmetry in 't Hooft's construction
\cite {tHooft2}.
On the other hand, if some $k$ number of 
eigenvalues $r^m_i$ coincide, the 
unbroken symmetry group becomes  $SU(k)\times  U(1)^{N-k-1}$. 
In fact, for all coincident $r^m_i$'s the symmetry group 
is restored back to $SU(N)$\footnote{Here we are a bit
loose on the difference between $U(N)$ and $SU(N)$.
In $SU(N)$ all the eigenvalues cannot coincide unless 
they are all equal to zero. Otherwise, only $N-1$
eigenvalues can be identical. Here we can consider $U(N)$ instead of
$SU(N)$. Then we have $N$ independent eigenvalues all of which can be 
identical. One of these eigenvalues, however, is due to the extra $U(1)$,
and corresponds to the overall
center-of-mass motion of the zerobranes, which can be dropped - see below.}.
We will discuss physical implications of this fact in 
the next section. Note that 
the expression (\ref {Phivacuum})  
is the static solution of the
equation of motion of (\ref {MM}):
\begin{eqnarray}
{\ddot X_i}^a ={2\over (2\pi \alpha')^2} ~
 \Big ( \Big [ X_j~X_i \Big ] X_j \Big )^a~.  
\label{EOM}
\end{eqnarray} 
An important issue is to specify symmetries of the
ground state defined by the solution
$X_i^{\rm cl}~=~{\rm diag}
(r_i^1, r_i^2,...,r_i^N)$. Note that $\sum_{k=1}^{N}~r_i^k=0$.
Thus, there are $N-1$ independent values of $r^m_i$'s for each 
spatial component $i=1,2,3$. Since these  
constants can be arbitrary, the solution (\ref {Phivacuum})
takes values in the space ${\bf R}^{3(N-1)}$. Moreover, 
the solution (\ref {Phivacuum}) is invariant under
permutations of $r^m_i$'s. These permutations form a group 
$S_N$,  which is the Weyl subgroup of $SU(N)$. Thus, the space 
\begin{eqnarray}
{{\bf R}^{3(N-1)}\over S_N}
\label{SN}
\end{eqnarray}
can be identified with the moduli space 
of $N-1$ electrically neutral point-like
objects. Let us tentatively call these objects QCD zerobranes
and study their properties. 
\vspace{0.3in} \\
{\it  3.1. Zerobrane interactions}
\vspace{0.1in} 

In this subsection we study interactions between QCD zerobranes.
For simplicity we will concentrate on the $U(2)$ gauge group.
There should be two zerobranes in this case.
Below we are going to show that the potential between them
is linearly rising with distance.
Hence, the only non-interacting configuration will be 
produced by the zerobranes  which are placed on top of each other. 
Some physical implications of this fact will be 
discussed at the end of this section. First, let us
calculate the potential between the zerobranes\footnote[3]
{Later we would like to identify zerobranes with QCD monopoles.
Magnetic charges  of monopoles for $U(N)$ group take the values:
$(1,-1,0,...,0),~(0,1,-1,0,..,0),\dots$. These charges 
cannot be manifestly seen in
the matrix model Lagrangian since there are no local $U(1)$ gauge fields 
there, and vector fields are expected to be confined in flux tubes. 
Nevertheless, these charges can be understood
by assigning  Chan-Paton type factors to  
the flux tubes stretched between $N-1$ zerobranes.}. 

Let us make a standard decomposition of the $X_i$ field
into its classical and quantum parts:
\begin{eqnarray}
X_i~=~X_i^{\rm cl}~+~\delta X_i~.
\label{decomposition}
\end{eqnarray}
Here, 
\begin{eqnarray}
X_1^{\rm cl}~=~{1\over 2}~\left (
\begin{tabular} {c c}
$r$ & 0 \\
0 &  $-r$ 
\end{tabular}
\right )~,~~~X^{\rm cl}_2=0~,~~~X^{\rm cl}_3=0~.\label{decomposition1}
\end{eqnarray}
That is, the two zerobranes  are at a distance $r$ apart from each other 
along one spatial direction.
The effective potential between  
point-like objects described by 
(\ref {MM}) has  been calculated 
in the context of M-theory zerobrane quantum mechanics
\cite {DKPS,BFSS} (for a review, see \cite {Taylor}). 
Thus, we can use the Matrix theory results \cite {BFSS} 
by keeping only the corresponding non-supersymmetric (that is, bosonic)
four-dimensional parts
(see \cite {DKPS,Taylor,ishibashi,Periwal}). 
In the path integral representation, after the appropriate gauge fixing,
one integrates out  quantum fluctuations $\delta X_i$.
As a result, one finds the effective potential (see, for instance, 
\cite {ishibashi,Periwal}):
$$
V_{\rm eff}~\propto ~{\rm ln}~{\rm det}~(~-\partial_t^2~+~r^2~)~,
$$
where we have temporarily set $2\pi \alpha^\prime=1$.
The finite, $r$ dependent part of this potential with 
the appropriate  dimensionality is given by:
\begin{eqnarray}
V_{\rm eff}~\propto ~{1\over  \alpha'}~r~.
\label{potential}
\end{eqnarray}
Hence, there is a linearly rising potential between QCD zerobranes.
\vspace{0.3in}\\ 
{\it 3.2. Compactification of the matrix model and the conjecture}
\vspace{0.2in}

Next, we would like to discuss the relation between the matrix model
(\ref {MM}) for zerobranes and pure QCD. For this let us
compactify the matrix model on a three-torus
$T_{R}={\widetilde S}^1\times {\widetilde S}^1 \times {\widetilde S}^1$. 
The detailed procedure for compactifying the matrix model on a torus is 
described in \cite{Taylor}, so our discussion here will be brief.
Instead of studying the motion of $N$ zerobranes on 
a torus $T_R \equiv ({\bf R} \times {\bf R} \times{\bf R})/
({\bf Z}\times {\bf Z}\times {\bf Z})$ 
one studies the motion of an infinite family of
zerobranes on  ${\bf R}^3$. 
Then one imposes a set of constraints which guarantee  
that  the system is invariant under the discrete group 
${\bf Z}\times {\bf Z}\times {\bf Z}$. The fields $X^i_{ma,nb}$
describing zerobranes (along with the $SU(N)$ indices $a,b=1,\dots,N$) 
now carry a pair of winding indices 
$m,n \in {\bf Z} $ (for a given value of the index $i=1,2,3$). 
Thus, the compactified matrix quantum mechanics
(\ref {MM}) transforms explicitly into the form (\ref {Philagrangian})
upon  the substitutions 
$$ X^i_{ma,nb}\leftrightarrow \Phi^i_{ma,nb}~,~~~g_s\leftrightarrow g~,$$ 
and the duality relations defined in section 2. 
As we have already mentioned, 
the winding modes in
the matrix model transform into the Kaluza-Klein momentum modes in QCD. 
The two theories compactified on dual tori are equivalent. 
That is, they are
T-dual to each other. Does this mean that the degrees of freedom of
one theory can be ``understood'' in terms  of the degrees of freedom
of the other one and {\em vise-versa}?
In the case of string theory D0-branes such a relation exists. 
The low energy limit of the D0-brane matrix quantum mechanics 
is described by pure super-YM. The analogous picture 
for gluons can be thought of in the nonsupersymmetric case too. 
Generically the symmetry of the dual theory 
(\ref {MM}) is $U(1)^{N-1}$. 
However, once some of the $X_i$  eigenvalues coincide, 
the symmetry group is enhanced. 
For instance, if all the $N-1$  
eigenvalues coincide  the symmetry is restored back to $SU(N)$.
Where are gluons in this picture? The condition that 
all the $N-1$ eigenvalues coincide defines the so called
${A_{N-1}}$ singularity of the $SU(N)$ group \cite {Arnold}. 
In this respect, 
$SU(N)$ {\it massless gluons} can be viewed as  massless states 
emerging at the singular points in 
the QCD zerobrane  moduli space where  the coordinates of all the 
different $N-1$ zerobranes  coincide.
It is attractive  to adopt the following
{\it illustrative} picture of this phenomenon. 
Since there is a linearly rising
potential between  QCD zerobranes, one can imagine   
a stretched flux tube, a sort of stretched string  
between these point-like objects.
Suppose this flux tube  has excitations with vector particle 
quantum numbers (albeit it is not clear how to see this explicitly).
When  the QCD zerobranes   are separated at some finite distance 
from  one another, 
the flux tube is stretched and its excitations are massive.
Masses of these ``gluons''  will be proportional to the 
separation between the corresponding QCD zerobranes
$$
m_{kl}~\propto ~{1\over \alpha'}~ |~r_k~-~r_l~|~. 
$$
However, when zerobranes  come closer to each other, 
the flux tube  relaxes. 
In fact, when the zerobranes  coincide, some of the  
massive string excitations might become massless
and fill in 
the  multiplet of massless gluons of 
$SU(N)$\footnote[1]{This is exactly what happens in superstring theory 
when branes sit on top of each other \cite {Witten}. Moreover, 
the picture of the stretched string is precise in this case
since a critical open string theory indeed has a massless 
vector particle in its spectrum. Nonetheless, in this case we are dealing
with the Higgs mechanism, whereas in the case of QCD monopoles gluons are
actually confined.}. 
What about zerobranes, could they be seen in terms of pure QCD?
In the 't Hooft's construction
there are nonperturbative monopoles in pure QCD 
which emerge  as singularities in the theory of 
gluons. That is to say, these monopoles are not
seen in pure QCD as solitons when the gauge group is unbroken.
However, they appear as certain singularities 
in the gauge fixing condition when 
the $SU(N)$ gauge theory is {\it projected} to a 
$U(1)^{N-1}$ theory by a special choice of gauge. 
We would like to conjecture that 
nonperturbative monopoles of QCD can  be described 
in the T-dual form by zerobranes of the matrix quantum mechanics (\ref {MM}).
Besides the symmetry breaking pattern, which is the same for monopoles and 
zerobranes,  
this conjecture is supported by the results of our work  \cite {det}.
In this work we studied pure QCD with the theta-term. It is known that
monopoles acquire electric charges due to the Witten effect
\cite {witten} once non-zero theta-angle is 
introduced \cite {tHooft2}. 
If the identification of zerobranes  
with monopoles is 
correct, then the Witten effect should also be seen for zerobranes 
within the matrix model (\ref {MM}) amended by the theta-term. In other words, 
if one calculates the interaction force between the point-like objects
of the matrix model with the theta-term, then 
this force should depend on the theta angle. 
We found that this is indeed the case, {\em i.e.}, the 
interaction potential between pure QCD zerobranes of (\ref {MM}) 
depends on the theta angle \cite {det}. 
This can be interpreted as the fact that 
zerobranes (monopoles) acquire  electric charges due to the Witten effect, 
and that these electric charges change the 
interaction force between them \cite {det}.
Note that in QCD monopoles emerge  as singularities in the theory of 
gluons. On the other hand, in the T-dual picture 
QCD monopoles are manifest but 
{\it massless gluons} emerge at certain singularities 
in the monopole moduli space.  
Adopting the conjecture
formulated above, we study condensation of QCD monopoles in the following section. 
\vspace{0.2in} \\
{\bf 4.  Monopole Condensation}
\vspace{0.2in}

In this section  we intend 
to study the limit $L>>\zeta$, that is, $R<< \zeta$. In this limit
matrix quantum mechanics (\ref {Philagrangian},\ref {MM})
is placed in a small volume, and it should describe
large volume YM theory. 
According to 
(\ref {couplings}), the coupling constant  
$g$ in the matrix quantum mechanics can be kept fixed at a small value
even if the YM coupling $g_{\rm YM}$ is large. 
Thus, in what follows we will study the regime where
$$ L~>>~\zeta~,~~~
R~<<~\zeta~,~~~ g_{\rm YM}^2~>>~1~,~~~g~<<~1~.
$$
However, in the matrix quantum mechanics we still have a large number of 
light winding modes which complicate the dynamics. 
Indeed, pure QCD in a large volume limit is 
a complicated theory containing not
only the zero-modes of the compactification but also a large number of
light Kaluza-Klein modes whose masses scale as $1/L\ll \Lambda_{\rm YM}$. 
The matrix quantum mechanics in a small volume is just {\it as complicated} 
as it
contains light {\em winding} modes (which map to the Kaluza-Klein modes in
the T-dual YM description) whose masses scale as $R/\alpha^\prime=1/L$.  
However, certain aspects of the YM theory in a large volume (which is a
strongly coupled theory) might be more transparent in the small volume
matrix quantum mechanics approach. 
We will consider below the limit when the monopoles are 
moving very slowly. That is to say, their momenta are much smaller than the 
masses of light winding modes. In this regime 
we expect that the winding modes
are not yet excited and certain properties can be simplified. 
In particular, this should be applicable for studying 
condensation of slowly moving
monopoles. 
In this approximation we can use the Lagrangian 
(\ref {Philagrangian}) where winding modes are neglected 
and derive condensation of 
QCD monopoles. The first thing to do is  
to write down the quantum mechanics of these massive objects in the 
monopole moduli space approximation \cite {Monton}\footnote{Recent 
discussions on  the monopole moduli space approximation
can be found in \cite{Weinberg} and references therein.}. 
That is, let us suppose that there is a  region in the 
monopole moduli space where interactions between heavy, 
slowly-moving monopoles are weak. 
This is the regime when monopoles are close to each other. 
Then the field $\Phi (t)$ in the 
moduli space approximation can be written as:
\begin{eqnarray}
\Phi_i(t)~=~U~\Phi^{\rm cl}_i (t)~U^{+}~=~
U~{\rm diag} \Big (~ r_i^1(t),~ r_i^2(t),...,r_i^N(t)~\Big )~U^+ ~,
\label{monopoles}
\end{eqnarray}    
where the constant moduli have now become 
time dependent quantities
$r_i^m(t)$ \cite {Monton}. As we stressed above, they define positions of
point-like QCD monopoles. 
Notice that  in the moduli space approximation defined by (\ref
{monopoles}) the commutator term in (\ref {Philagrangian}) vanishes.
Thus, the model (\ref {Philagrangian}) reduces to 
a quantum mechanics of heavy, slowly-moving QCD monopoles 
with the  following simple Hamiltonian:
\begin{eqnarray}
H~=~{g~\sqrt {\alpha'}}~\sum_{m=1}^{N-1} 
\Pi^m_i(t) \Pi_i^m(t)~.
\label{Hmonopoles}
\end{eqnarray}
Here $\Pi_i^m$ stands for the conjugate momentum in the $i$'th spatial 
direction of  the $m$'th QCD monopole\footnote{In this Hamiltonian
we have dropped the term describing the center-of-mass motion of 
the system of $N-1$ monopoles.}. 
The mass of a  QCD monopole  is therefore given by:
\begin{eqnarray}
M~=~{1\over 2 g \sqrt {\alpha'}}~=~{\Lambda_{\rm YM}\over 2 g}~.
\label{mass}
\end{eqnarray}
As a result of (\ref {period}) the monopole coordinates
are periodic variables. Therefore,
the Hamiltonian (\ref {Hmonopoles}) has the unique normalizable 
eigenstate with zero momentum:
\begin{eqnarray}
\Psi_0~(r_i^1,r_i^2,..,r_i^N)~=~ {1\over (2\pi R)^{3/2}}~.
\label{eigenstate}
\end{eqnarray}
This state satisfies the Gauss's law constraint (\ref {gauss1}).
In fact, it describes monopole condensation
in the moduli space approximation\footnote{The eigenvalue for this 
eigenstate is  zero in the non-relativistic approximation we 
deal with. In the relativistic case the non-zero eigenvalue 
would be given by the QCD monopole  mass (\ref {mass}).}.
Thus,  the QCD monopoles can condense.
Notice that if $R\rightarrow 0$ the wavefunction 
(\ref {eigenstate}) and the probability density
for the condensate blow up. However, 
the probability itself $\int_0^ {2\pi R} |\Psi_0|^2 d^3x$
equals to 1. Moreover, a state with zero momentum 
is expected to have a constant wavefunction, like the one in  (\ref
{eigenstate}), in accordance with the uncertainty principle.    

The above discussion is 
valid in the approximation of slowly-moving,
almost non-interacting monopoles
(the moduli space approximation of 
(\ref {monopoles})). 
For generic values of $\Phi_i(t)$
the commutator in the potential (\ref {MM})
is non-zero. Hence, matrices $X_i$ cannot be diagonalized
simultaneously. Nevertheless, 
monopole positions can still be defined as eigenvalues of 
$X_i$ \cite {Witten}. The off-diagonal elements then describe
interactions between monopoles.

In the previous section we 
have seen  that the potential between the monopoles 
is linearly rising. Therefore, the approximation adopted in this section 
is valid only if monopoles are extremely close to each other,
or, more precisely, when they sit on top of each other. 
This means that they can condense only as 
some composites.
Moreover, in the state of lowest energy 
the constituents in a condensate should sit on top each other\footnote{
A condensed monopole-antimonopole pair in some sense resembles
a condensed quark-antiquark pair in QCD: there is a linearly rising
potential between quarks, yet they  can be considered as free particles 
when they come close to  each other, and, finally, they condense in
quark-antiquark pairs in the S-wave channel. All these
properties seem to be shared by the QCD monopoles as well.}.
Once the pair is excited in the condensate, 
monopole-antimonopole constituents will start
to move with respect to each other with opposite momenta, keeping the
total momentum equal to zero. In this case there is a linearly rising
potential between these  objects (\ref {potential}). 
Finite velocity corrections can also be calculated to this 
potential using the results of \cite {Periwal}\footnote{
The phenomenon of monopole-antimonopole condensation
is S-dual to Cooper's phenomenon
of ``electric charge" condensation.  
However, the  potential between constituents in a condensed 
monopole-antimonopole pair is linearly raising.  
This differs from Cooper's condensation
which gives rise to a weakly coupled pair.
Strongly coupled monopole-antimonopole condensation might be suggestive 
of a possible relation to high-temperature superconductivity where we are
dealing with the strong coupling regime.
That is, the QCD ground state might be ``electric-magnetic''
dual to the ground state of a theory describing 
high-temperature superconductivity.}.
\vspace{0.3in} \\
{\bf 5. Large $N$  QCD {\em vs.} Bosonic Membrane Theory}
\vspace{0.1in}

In this section, using the results discussed above, we 
argue that in a certain regime large $N$ QCD 
can be  described by  a  {\it five} dimensional theory of a closed
bosonic membrane. 

Thus, consider the regime 
$L<<\zeta$. That is, pure QCD is compactified on a small 
three-torus. As we discussed in section 2, 
in the T-dual picture 
this theory is described by the large volume $R>>\zeta$ matrix 
quantum mechanics 
(\ref {Philagrangian}). Thus, to calculate the spectrum of pure 
YM in the small volume limit one can use the matrix quantum mechanics
(\ref {Philagrangian}) and study the corresponding 
Schr\"odinger equation in the large volume limit \cite {Luscher}. 
Furthermore, a remarkable discovery was made in 
\cite {Hoppe1,Hoppe2} where the matrix quantum
mechanics (\ref {MM})
in the large $N$ limit was shown to be equivalent to a light-cone
theory of a closed bosonic membrane.
As a consequence, the membrane theory was used in  \cite{gabad} to study 
the small volume YM spectrum in the large $N$ limit.

In this section we will
briefly discuss  the relation between the membrane theory, large $N$ 
matrix quantum mechanics and YM model from the point of view
of T-duality. In particular, let us review how 
the membrane action reduces  to the matrix model 
(\ref {Philagrangian}, \ref {MM}).
We start with the theory of a closed bosonic membrane in 
five dimensional space-time.
We are going to present  the basic features  of 
the membrane Hamiltonian construction in the light-cone gauge. 
For details the reader is  referred 
to the original papers \cite {Hoppe1,Hoppe2}\footnote[2]{
For discussions on the relation between large $N$ gauge groups and
toric surfaces, see \cite {Zachos}.}. 

The membrane action in flat Minkowski
space-time  can be written as follows:
\begin{eqnarray}
S_m=-T \int d^3 \sigma \sqrt{|{\rm det} g_{\alpha\beta}|}~,
\end{eqnarray}
where $T$ is the membrane tension with the dimensionality  
of mass cubed; $\sigma_\alpha$, $\alpha=0,1,2$,
are the coordinates on the membrane world-volume;
$g_{\alpha\beta}$ denote the components of  the induced metric in the 
membrane world-volume
\begin{eqnarray}
g_{\alpha\beta}(\sigma)\equiv {\partial X^\mu (\sigma )  
\over \partial \sigma^\alpha}
{\partial X_\mu (\sigma)  \over \partial \sigma^\beta}~,
\end{eqnarray}
where $X_\mu$, $\mu=0,1,2,3,4$, are the target space-time coordinates.

The membrane action is reparametrization invariant.  
Hence, not all of the variables in the action are independent.
One should carry out the gauge fixing procedure. 
It is convenient to introduce the light-cone coordinates:
\begin{eqnarray}
X^{\pm}={1\over \sqrt 2} (X^4\pm X^0)~, \nonumber
\end{eqnarray}
and choose the light-cone gauge
\begin{eqnarray}
X^{+}(\sigma)=X^{+}(0)+\sigma_0~. \nonumber
\end{eqnarray}
The light-cone gauge does not completely fix the 
gauge freedom of the membrane action.
As a result, there still is 
a residual local  invariance left.  
Hence, one expects to have
the Hamiltonian of the theory accompanied by a  constraint equation.
The detailed discussion and the construction 
of the Hamiltonian is given in 
\cite {Hoppe1,Hoppe2}. Here we present the final result.
The expressions for the mass squared operator and the constraint 
can be written as follows: 
\begin{eqnarray}
{{\hat M}^2\over 2}= \left [ {1\over 2}
{\cal P}^a_i{\cal P}^a_i+ 
{T^2 \over 4}(f^{abc}X_i^bX_j^c)^2 \right ]~,
\label{membrane}
\end{eqnarray}
\begin{eqnarray}
{f} ^{abc}X_i^b{\cal P}_i^c=0~,~~~i,j=1,2,3~. \label{memconstraint}
\end{eqnarray}
Here $f^{abc}$ stand for the corresponding structure constants which we
will specify in a moment.
The canonical coordinates and conjugate momenta are the functions 
of the time variable $\sigma_0$ only. 
The coordinates $X_i^a$ in this expression 
are the coefficients of the harmonic expansion 
of the space-time coordinates $X_i$ on the surface of the membrane.
For instance,  if the membrane has  the 
topology of a two-sphere or a two-torus, then
the harmonic expansion mentioned above 
is just an  expansion of the space-time coordinates 
in the basis of the harmonic functions
on the corresponding surface: 
$$
X_i(\sigma)=\sum _{a=1}^{\infty} X_i^a(\sigma_0) Y^a(\sigma_1,\sigma_2 )~,
$$
where $Y^a(\sigma)$'s are the harmonic functions.
Here the harmonic functions $Y^a(\sigma)$ form 
a representation of the Lie algebra
of the $SU(\infty)$ gauge group 
\cite {Hoppe1,Hoppe2}\footnote[8]{The $SU(\infty)$ 
group (and its Lie algebra) should be understood as the $N\rightarrow \infty$  
limit of the $SU(N)$ group.}. The structure constants $f^{abc}$ are those
of $SU(\infty)$.
Thus, the  $SU(\infty)$ gauge group appears due to the 
reparametrization invariance of the membrane 
action.

The expression (\ref {membrane}) 
coincides, up to some rescalings, with the Hamiltonian
which can be derived from (\ref {Philagrangian}) in the 
$N \rightarrow \infty$ limit. 
Indeed, the Hamiltonian corresponding to (\ref {Philagrangian})
reads:
\begin{eqnarray}
{1\over 2 g \sqrt{\alpha'} }~H~=~{1\over 2} ~\Pi_i^a\Pi_i^a~+~
{1\over (2 g \sqrt{\alpha'})^2 }~{1\over 4 (2\pi \alpha')^2}~\Big (~
f^{abc} \Phi_i^b\Phi_j^c ~ \Big )^2~.
\label{hamham}
\end{eqnarray}
The constraint equations in (\ref {gauss1}) 
and (\ref {memconstraint})
(which correspond to the Gauss's law (\ref {gauss})) 
are also identical.
Thus, the matrix quantum mechanics 
(\ref {Philagrangian}) in the large $N$ limit
is equivalent to the theory of a closed bosonic membrane
with the topology of a sphere or a torus \cite {Hoppe1,Hoppe2}.
On the other hand, we have argued above that the matrix
quantum mechanics (\ref {Philagrangian}, \ref {MM})
compactified on the three-torus $T_R$ describes
QCD monopoles and pure QCD on a dual torus $T_L$.
Thus, a five dimensional theory of a bosonic membrane
is a T-dual description of pure QCD. QCD point-like
monopoles in this case can be regarded as 
constituents of the membrane.
The formal transition between the two systems is via the following
substitutions:
\begin{eqnarray}
{\hat M}^2~\leftrightarrow~ {1\over g \sqrt{\alpha'}}~H~,~~~~~
T^2~\leftrightarrow ~ {1\over (4\pi g \alpha' \sqrt {\alpha'})^2}~.
\label{aketiket}
\end{eqnarray}
Thus, one can calculate the membrane spectrum in terms of 
$M$ and $T$, and then find the corresponding YM spectrum 
(in the regime where YM is compactified on a small torus and is weakly
coupled) using
(\ref {aketiket}) \cite {gabad}.
\vspace{0.3in} \\
{\bf 6. Speculations}
\vspace{0.1in} 

The discussions in the previous section lead us to speculate on a
possibility of an underlying {\em five} dimensional theory which might
provide an adequate low energy description of strongly coupled 
pure QCD. In particular, one could notice 
an obvious analogy between the discussions in the
previous sections and what happens in the context of M-theory. Thus,
the strong coupling limit of 10 dimensional 
Type IIA is believed to be described
by 11 dimensional M-theory whose low energy effective theory is 11
dimensional supergravity. On the other hand, the large $N$ quantum mechanics
of Type IIA D0-branes is believed to describe M-theory in the infinite
momentum frame \cite{BFSS}. The former also arises upon the light-cone
quantization of the M2-brane \cite {Hoppe2}. 

The analogy with QCD monopoles is then clear. The latter are
analogous to D0-branes, and Type IIA should be analogous to the QCD string
theory, which is believed to be non-critical (unlike Type IIA)
\cite{Polyakov}. Because of this property it is likely that this string
theory in the strong coupling limit 
is intrinsically 5-dimensional, and, perhaps, the string expansion
may not even be adequate so that some sort of ``membrane expansion'' (as in
M-theory) might be a more appropriate description. If so, it is tempting
to conjecture that there is an analog of M-theory, which is a 5-dimensional
theory (let us tentatively call it ``${Q}$-theory''), 
whose low energy effective field theory has
solitonic membrane solutions. These membranes are made of 't Hooft's QCD
monopoles. The matrix quantum mechanics (once we also include
non-perturbative effects and take the large $N$ limit) describing these 
monopoles then might also describe some other properties of large $N$
QCD. In particular, certain regimes where we might not naively expect the
matrix quantum mechanics to be applicable could still be adequately
described by the latter in analogy with the M-theory case. One possible
indication of this might be the fact that the glueball spectrum computed in
\cite{gabad} using the membrane theory as the starting point is in a good
agreement with the lattice data. In fact, in this case one can argue from
the QCD viewpoint that such an agreement is not an accident as the size
of the corresponding glueballs is much smaller than the QCD correlation
length $\zeta=1/\Lambda_{\rm YM}$. 

If such a five dimensional ${Q}$-theory indeed exists, this would have
interesting implications for QCD. First, the matrix quantum mechanics would
then provide a computational tool for QCD. Certain sectors of the glueball
spectrum might also be computable by quantizing the membrane. The magnetic
dual of the letter (in 4+1 dimensions) is an instanton. It is tempting to
identify these instantons with the QCD instantons which would be consistent
with the expectation that the latter are also made of point-like QCD 
monopoles \cite {Lee}. 

At present it is not completely clear whether the above conjectures, which
are based on the analogy with M-theory, will hold. In particular, here we
do not expect any non-renormalization theorems as we have no supersymmetry.
Generalization to supersymmetric QCD might therefore be desirable as in the
latter case supersymmetry might yield some simplifications (albeit there might
also be various possible complications). These and other issues are
currently under investigation.      
\vspace{0.3in} \\
{\bf Acknowledgments}
\vspace{0.1in} 

The authors are grateful to Gia Dvali for useful discussions.
The work of G.G. was  supported by the grant
NSF PHY-94-23002. The work of Z.K. was supported in part by the 
grant NSF PHY-96-02074, and the DOE 1994 OJI award. Z.K. would also 
like to thank Albert and Ribena Yu for financial support.

\end{document}